\documentclass{cpp2e}
\usepackage{epsfig, floatflt, rotating, latexsym, amssymb}
\shorttitle{Plasmon dispersion of a nonideal plasma} %Please fill in.
\shortauthor{V.~Golubnychiy et al.} %Please fill in.
\contvolum{} %Dont't fill in.
\contyear{2001} %Dont't fill in.
\contnumber{} %Dont't fill in.
\contpages{} %Dont't fill in.
\title{
Plasmon dispersion of a weakly degenerate nonideal one-component plasma
}
\author{V.~Golubnychiy$^1$,
M.~Bonitz$^1$, 
D.~Kremp$^1$, and M.~Schlanges$^2$ }
\address{
$^1$Fachbereich Physik, Universit{\"a}t Rostock,
Universit{\"a}tsplatz 3, D-18051 Rostock \\
$^2$Institut f\"ur Physik der Universit{\"a}t Greifswald,
Domstrasse 10a, D-17489 Greifswald  \\
e-mail:vova@mars.mpg.uni-rostock.de}
\begin{document}
\setcounter{page}{1}
\makeheadings
\maketitle
%-------------------------------------------------------------------
\begin{abstract}
Classical Molecular Dynamics simulations (MD) for a one-component
weakly degenerate plasma are presented. Using an effective quantum pair potential 
(Kelbg potential), the dynamic structure factor and the dispersion of Langmuir 
waves are computed. 
The influence of the coupling strength
 $\Gamma$ and degree of degeneracy $\rho \Lambda^3$ on these properties is 
discussed. The results are compared with predictions of mean-field theories. \\
\end{abstract}
%------------------------------------------------------------------
%\pacs{52.25 Kn, 52.65.-y, 52.65.Pp}
%PACS: 52.25 Kn, 52.65.-y, 52.65.Pp \\
%-------------------------------------------------------------------

%-------------------------------------------------------------------
%\newpage
The dielectric properties and the oscillation spectrum of charged particle systems 
have been studied in great detail for many decades, e.g. \cite{i92,clwh82,env99}. 
i). In the case of {\em weakly correlated 
classical} plasmas, where the coupling parameter 
$\Gamma=(4\pi \rho/3)^{1/3}e^2/4\pi \epsilon_0 k_BT$ ($\rho$ is the electron 
density) is small, the appropriate description is given by the Vlasov theory. ii) On the 
other hand, {\em weakly correlated quantum} plasmas with $\rho \Lambda^3 > 1$, ($\Lambda$ is the 
DeBroglie wave length), are well described by the random phase approximation (RPA).
These two mean-field theories neglect correlation effects, in particular they do 
not take into account plasmon damping due to collisions. The latter effect becomes 
essential for increasing coupling, $\Gamma>1$. iii) {\em Strong coupling in classical} 
plasmas can be efficiently studied using molecular dynamics simulations, 
e.g. \cite{clwh82,env99}. However, this powerful method is not applicable to quantum 
systems. iv) Alternative approaches to the dielectric properties of quantum systems, 
such as density functional or quantum kinetic theory, e.g. \cite{i92,kwong-etal.00prl}, 
are usually successful up to moderate coupling, $r_s\equiv{\bar r}/a_B<1$, where 
${\bar r}$ and $a_B$ denote the mean inter-particle distance and the Bohr radius, 
respectively.

This brief survey shows that plasmas in the region of {\em moderate coupling and degeneracy} 
are difficult to describe reliably by means of established methods. In this paper, we analyze 
the dynamic structure factor and the Langmuir plasmon dispersion for weakly degenerate 
plasmas, $\rho \Lambda^3 \le 1$, at intermediate coupling, $\Gamma\le 4$.
For these parameters, it is possible to perform classical MD simulations
using effective quantum pair potentials. Such potentials have been rigorously derived from the 
two-particle Slater sum using Morita's method by Kelbg and co-workers \cite{kelbg}. In 
this paper we alternatively use the Kelbg potential and the Coulomb potential which allows us 
to study the influence of quantum diffraction effects on the dynamic properties of an OCP.
We find that quantum effects tend to soften the wave vector dispersion.

The dielectric and dynamic properties of an N-particle system can be 
derived from the density-density correlation function $A(\vec{k},t)$,
which is defined as \cite{i92}: 
\begin{eqnarray}
A(\vec{k},t) = \frac{1}{N} \langle\rho_{\vec{k}}(t) \rho_{-\vec{k}}(0)\rangle,
 \qquad
\rho_{\vec{k}}(t) = \sum_{i=1}^{N}  e^{i \vec{k} \vec{r_i}(t)}.
\label{rho_k}
\end{eqnarray}
Here, $\rho_{\vec{k}}(t)$ is the spatial Fourier component of the density
which is computed from the trajectories $\vec{r}_i(t)$ of all particles. 
The dynamical structure factor is just the Fourier transformation of the density-density 
correlation function
\begin{eqnarray}
S(\vec{k},\omega)= \frac{1}{2 \pi} \int\limits_{-\infty}^{+\infty} dt \, 
e^{i\omega t} \, A(\vec{k},t). 
\label{s_md}
\end{eqnarray}
%\vspace{-0.1cm}
The trajectories of particles $\vec{r}_i(t)$ can be directly obtained by numerical solving 
classical (Newton's) equations of motion for the interacting N-body system which include 
all binary Coulomb forces, e.g. 
\cite{clwh82}. To include quantum diffraction effects, one can replace the Coulomb potential 
by an effective quantum potential which was derived by Kelbg \cite{kelbg}
\begin{eqnarray}
U_{\rm KELBG}(r,T)=4\pi e^2 \!\left(\frac{1-\exp(-r^2/\lambda^2)}
{r}+\frac{\sqrt\pi}{\lambda}{\rm erfc}(r/\lambda)\right)
\label{kelbg}, \qquad \lambda = \frac{\Lambda}{\sqrt{2 \pi}}.
\end{eqnarray}
\begin{floatingfigure}[r]{9cm}
\mbox{	\epsfig{file=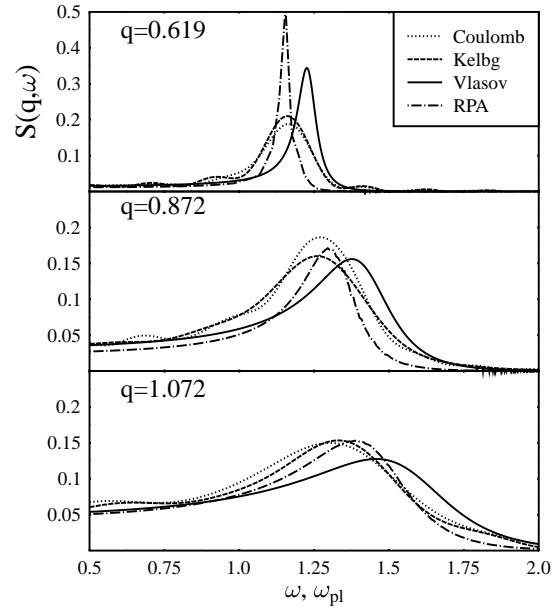, width=9cm}}
\vspace{-4cm}
	\caption{\small{
Dynamical structure factor an OCP at $\Gamma = 1$ and $\rho \Lambda^3 = 0.1$
from MD simulation with Coulomb and Kelbg potential.}
\vspace{0.2cm}}  
\label{fig1}
\end{floatingfigure}
The Kelbg potential (\ref{kelbg}) is finite at zero distance, correctly accounting for the 
Heisenberg uncertainty. Most importantly, it is the exact quantum potential in the case of 
small $\Gamma$ and we expect that, also at moderate coupling, it accounts for the dominant quantum 
effects. 
This potential is used in the MD simulations shown below. Furthermore,
 the long range part of the interaction was computed in standard way by using Ewald 
summation procedure and 3-dimensional lookup tables, for details cf. \cite{md_ocp}.                
 
 We have performed a series of simulations for varying values of $\Gamma$ and $\rho \Lambda^3$,
using the Coulomb and the Kelbg potential, respectively. 
Fig.~1. shows the dynamical structure factors $S(q,\omega)$, where
$q$ is the dimensionless wave number, $q=|{\bf k}|{\bar r}$. In the figure, also the results for Vlasov and RPA theories are 
plotted. The structure factor is clearly peaked at the frequency of the optical (Langmuir) 
plasmon. The width of the peak, i.e. the damping of the oscillations, increases rapidly with 
growing wave number. As expected, the mean-field results have narrower plasmon peaks 
which is due to the neglect of inter-particle collisions - their widths is solely due to 
collisionless Landau damping.
In contrast, the MD results fully include Landau and collisional damping. Further, as 
a result of the increased damping, the plasmon peak obtained in the MD simulations is 
shifted to lower frequencies.
Comparison of the simulation results with Coulomb and Kelbg potential 
shows only slight differences which are most pronounced at large wave numbers. 
The reason is that the interaction potentials differ only at small distances. 

\begin{floatingfigure}[r]{10.cm}
	\mbox{\epsfig{file=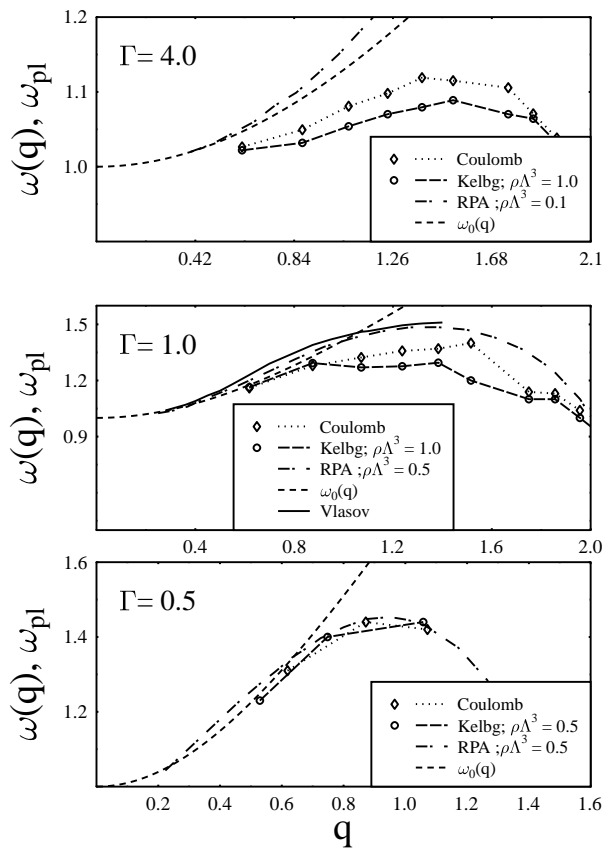, width=9cm}}
\vspace{-0.3cm}
	\caption{\small
	{Wave number dispersion of Langmuir waves for various coupling and degeneracy 
parameter from MD simulations with the Coulomb and Kelbg potentials. Furthermore, 
results from the Vlasov and RPA approximations are shown. $\omega_0(q)$ denotes 
the well-known analytical low $q$ limit of the Vlasov theory, see text.}
\vspace{0.2cm}}
\label{fig2}
\end{floatingfigure}

Increasing the degeneracy, we observe that the peak of the structure factor moves 
towards lower frequencies. A more quantitative comparison can be made by following the plasmon peak  
when the wave number is varied. The resulting wave number dispersion of Langmuir oscillations 
is shown in Fig.~2 for a dense OCP with three values of the coupling strength.
In the limit of small wave numbers, all models tend to the same frequency, the plasma
frequency $\omega_{pl}$. (Notice that simulation results are available only for finite 
wavenumbers, $q\ge q_{min}$, where $q_{min}$ is limited by the size of the simulation box.)
With increasing wave number, the frequency is expected to grow.
In particular, the Vlasov theory predicts, for small wave numbers, $k r_D < 1$, the 
behavior $ \omega_0(q)=\omega_{pl} \sqrt{1 + q^2/\Gamma}$. Our results show 
that, for $\Gamma=0.5$ and $q\le 0.7$ this formula works satisfactorily. However, 
for larger wave numbers, the plasmon frequency grows more slowly until it reaches 
a maximum around $q\approx 1$. For $\Gamma=0.5$, the plasmon dispersion is 
overall well reproduced by the RPA result which we found to be significantly more 
accurate than the Vlasov theory.

Let us now consider larger values of the coupling parameter $\Gamma$, cf. the upper two 
figure parts. Here, the differences between the mean field theories and the simulations 
are becoming more prononounced: the MD results yield an essentially lower plasmon frequency than 
the RPA and the Vlasov theory. Further, with increasing coupling strength the dispersion curves become more 
flat and their maxima decrease. 
Finally, let us analyze the influence of quantum
effects by comparing the simulations with the Coulomb and Kelbg potentials: we generally 
find that the dispersion curves with the Kelbg potential are below
those for the Coulomb case. This difference grows with increasing plasma degeneracy at 
constant $\Gamma$ confirming that this is a quantum diffraction effect.

%\newpage

\begin{floatingfigure}[r]{9.cm}
	\mbox{\epsfig{file=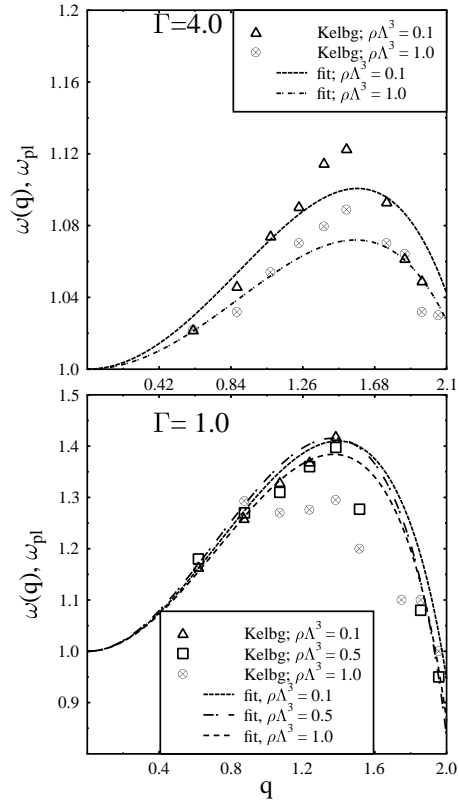, width=8cm}}
\vspace{-0.2cm}
	\caption{\small{Optical dispersion relations for different coupling and 
degeneracy parameter for the OCP with Kelbg potential. Symbols are MD results, 
lines the best fits.}}
\label{fig3}
\end{floatingfigure}

Finally, in Fig.~3 we study the dependence of the dispersion relations for the Kelbg 
potential on the degeneracy parameter $\rho \Lambda^3$ for two values of the coupling 
parameter. Again we confirm that increasing degeneracy
leads to a lowering (softening) of the plasmon dispersion. One may attempt to construct 
a fit for the plasmon dispersion which improves the above approximation 
$\omega_0(q)$. Fig.~3 shows the best fit constructed according to the ansatz
$ \omega(q)/\omega_{pl} = (1 + a q^2 + b q^4)^{1/2}$, where the points with smallest 
wave vector were given the largest weight \cite{fit}. The fit parameters for two 
values of $\Gamma$ and various degeneracies are 
summarized in the table. 

In summary, we have presented classical MD simulations with the Kelbg potential 
which allowed us to compute the optical plasmon dispersion of a weakly degenerate 
moderately correlated one-component plasma. The main conclusion is that increase 
of coupling and degeneracy influence the dispersion in similar way: they reduce
the plasmon frequency. The physical reason is simple: with increasing $\Gamma$, 
the Coulomb interaction is screened more and more. Thus, the binary interaction is
dominated by short-range effects which tends to soften the collective mode. 
Similarly, increase of the degeneracy leads to 
a growth of the electron wave length which, again, increases the efficiency of 
short-range binary interaction.

\newpage

\begin{acknowledgements}
This work is supported by the DFG (Schwerpunkt ``Laserfelder'') and a grant for 
CPU time at the NIC J\"ulich.
\end{acknowledgements}

\begin{table}
\caption{Fit parameters of the Langmuir dispersion curves shown on Fig.~\ref{fig3}. 
The fit equation was taken in the form 
$ \omega(q)/\omega_{pl} = (1 + a q^2 + b q^4)^{1/2}$.  $q$ is in units of ${\bar r}$.
The fit parameters for $\Gamma$ = 1 and $\rho \Lambda^3$ = 0.1 are less reliable, 
because of the absence of data for big wave vectors.
}
\begin{tabular}{|c|c|c|c|}
\hline 
$\Gamma$ & $\rho\Lambda^3$ & a & b \\
\hline 
1.0&0.1& 1.013 $\pm$ 0.031  & -0.260 $\pm$ 0.023\\ 
1.0&0.5& 1.074 $\pm$ 0.041  & -0.288 $\pm$ 0.013\\
1.0&1.0& 0.975 $\pm$ 0.055  & -0.259 $\pm$ 0.018\\
\hline
4.0&0.1& 0.169 $\pm$ 0.015  & -0.034 $\pm$ 0.006\\
4.0&1.0& 0.121 $\pm$ 0.007  & -0.025 $\pm$ 0.003\\
\hline 
\end{tabular}
\end{table}

%\newpage

%-------------------------------------------------------------------------------

\begin{thebibliography}{99}

\bibitem{i92} For an overview, see e.g. 
S.~Ichimaru, ``Statistical Plasma Physics'' Vol.{\bf I}, Addison-Wesley 
Publishing Company, 1992

\bibitem{clwh82} J.M.~Caillol, D.~Levesque, J.J.~Weis, and J.P.~Hansen, 
J. Stat. Phys {\bf 28}, 325 (1982)

\bibitem{env99} W.~Ebeling, G.E.~Norman, A.A.~Valuev, and I.A.~Valuev, 
Contr. Plasma Phys. {\bf 39}, 61 (1999)

\bibitem{kwong-etal.00prl} N.~Kwong, and M.~Bonitz, Phys. Rev. Lett. {\bf 84}, 1768 (2000)

\bibitem{kelbg} G.~Kelbg, Ann. Physik, {\bf 12}, 219 (1963);
{\bf 13}, 354; {\bf 14}, 394 (1964).

\bibitem{md_ocp} V.~Golubnichiy, M.~Bonitz, D.~Kremp, and M.~Schlanges,
subm. to Phys. Rev. E

\bibitem{fit}  Obviously, the wave vector dispersion of the simulation data is 
more complex than this formula. But to further refine the ansatz, more simulation 
data would be required. To improve the present fit, the      
MD-data for various $q$ were weighted by the values of the plasmon damping 
for that  $q$-value. 

\end{thebibliography}
\end{document}